# Measurement of Turbulence Injection Scale Down to the Chromosphere

**Authors:** Zac Bailey[1], Riddhi Bandyopadhyay[2], Shadia Habbal[1*], Miloslav Druckmüller[3]

**Affiliations:**

[1]Institute for Astronomy, University of Hawaiʻi; Honolulu, HI 96822, USA.

[2]Department of Astrophysical Sciences, Princeton University; Princeton, NJ 08544, USA.

[3]Faculty of Mechanical Engineering, Brno University of Technology; Brno, 616 69, Czech Republic.

*Corresponding author. Email: habbal@hawaii.edu

**Abstract**
The solar atmosphere displays a sharp temperature gradient, starting from spicules in the chromosphere at $2\times10^4$ K, outward into the corona exceeding $10^6$ K. Plasma turbulence produced by the transverse motion of magnetic fields anchored in the photosphere is likely the energy source producing this gradient. However, very little is known about the turbulent structures near the solar surface. Using the highest spatial resolution white-light total solar eclipse image to date, we measure the transverse correlation length at distances ranging from 0.33 to 9 Mm above the solar surface — two orders of magnitude closer than previous estimates. Our results show that the turbulence injection scale in the chromosphere is ~1.5 Mm, which we associate with the size of granules since they are the only structured features of comparable size. Further, the change in perpendicular correlation length with distance from the solar surface exhibits a plateau in the first 4 Mm, followed by a rapid increase until 9 Mm where it becomes shallower thereafter. We associate this radial gradient with the expansion of the magnetic field in the transition region between the chromosphere and the corona.

# 1. Introduction

Spectroscopic observations of the solar atmosphere during the 1868 total solar eclipse (TSE) identified the bright reddish band around the Sun as Hα 656.3 nm emission at $2 \times 10^4$ K, which Lockyer (1868) named the chromosphere. This band was characterized by highly variable jet-like structures, known as spicules (Athay & Holzer 1982). Also emitting in Hα were larger and far more complex structures called prominences as they 'protruded' above the occulted solar disk sometimes out to 0.5 $R_\odot$ (Lüst & Zirin 1960). A year later during the 1869 total solar eclipse, using a slitless spectrograph, Young (1870) and Harkness (1869) independently discovered a 'green' line at 530.3 nm extending beyond the chromosphere. It was not until 70 years later that Grotrian (1939) and Edlén (1946) associated it with the Fe XIV forbidden transition with a peak ionization temperature of $1.8 \times 10^6$ K. As a consequence of this milestone discovery, Parker (1958) theorized that the solar atmosphere could not remain bound to the Sun but had to escape in the form of a subsonic-supersonic solar wind reaching several 100's km/s at 1 AU. The existence of the solar wind was subsequently corroborated by Mariner 2 in-situ measurements at Mercury's orbit (Neugebauer & Snyder 1966).

Also discovered in-situ were coherent disturbances in the form of Alfvénic fluctuations which were thought to be Alfvén waves of solar origin (Belcher & Davis 1971). Consequently, theories invoking turbulence were pursued in earnest to account for the million-degree corona (Hollweg 1986). Later, Alfvén waves were directly observed in the chromosphere and the inner corona, with sufficient amplitude to account for coronal heating and solar wind acceleration (see, e.g. De Pontieu et al. 2007, Tomczyk et al. 2007, McIntosh et al. 2011).

Early theories of coronal heating and solar wind acceleration focused mainly on the corona and did not account for the two orders of magnitude temperature rise above the chromosphere, known as the Transition Region (TR). This sharp temperature gradient was first characterized empirically by space-based extreme ultraviolet (EUV) observations of the corona (Mariska 1992). Coupled with model studies (Fontenla et al. 1993, Avrett & Loeser 2008), the height of the TR was estimated to be about 2000 km. Hollweg (1986) was the first to include the TR in turbulent-driven solar wind models. Subsequent studies of turbulence heating assumed that Alfvén waves propagate outwards from the photosphere and are then reflected back by density gradients in the corona, leading to nonlinear interactions which transfer energy from large scale to smaller scale eddies (Matthaeus et al. 1999, Adhikari et al. 2024).

Despite recent progress, the exact mechanisms responsible for coronal heating and acceleration remain elusive (see e.g., Golub & Pasachoff 2010, Judge & Ionson 2024). Space missions such as the Parker Solar Probe and Solar Orbiter are providing unprecedented insights into the near-Sun plasma environment; they have already led to novel insights into the heating and acceleration processes (e.g., Bowen et al. 2020, Bandyopadhyay 2023, Shankarappa et al. 2024). Nevertheless, a crucial step in understanding these processes is to characterize the properties of turbulence near the solar surface.

A key parameter in characterizing the turbulent magnetic field is the characteristic scale of the large eddies where energy is injected. This can be estimated by the correlation length (L), which is related to the rate of turbulent energy deposition and transport of energetic particles. The correlation length directly enters in the von Kármán theory of turbulent energy decay which estimates the rate of energy injection from large scale as $dE/dt \sim E^{3/2}/L$, with E as the energy computed from the fluctuations in the turbulent medium (Karman & Howarth 1938, Hossain et al. 1995, Bandyopadhyay 2018, 2019). Similarly, in calculating parallel and perpendicular mean free path of cosmic rays and other energetic particles, the turbulence correlation length is a key input (Bieber et al. 1994, Zank et al. 1998, Matthaeus et al. 2003, Ruffolo et al. 2012). In the presence of a strong magnetic field the turbulence becomes highly anisotropic (e.g., Shebalin et all 1982, Oughton et al, 1992), and in such regime the key parameter is the perpendicular correlation length $L_\perp$. This is indeed the case for the solar wind close to the Sun.

While measurements of $L_\perp$ in the near-Sun environment are starting to emerge (Bandyopadhyay et al. 2020, Cuesta et al. 2022), at present, there are very few observations that probe the inner corona and yield quantitative estimates of $L_\perp$. Using the Coronal Multi-channel Polarimeter (CoMP) and its upgraded version (UCoMP) for coronagraphic measurements, Sharma and Morton (2023) were the first to infer $L_\perp$ at 0.2 $R_\odot$ (solar radii). However, estimations of turbulence correlation scale in the chromosphere and the TR are still absent. In this study we present the first measurements of $L_\perp$ over a distance range of 0.33 to 94 Mm, thus setting a new benchmark.

## 2. Total Solar Eclipse Observations and the Inference of $L_\perp$

The only remote observations that probe the solar atmosphere down to the surface are total solar eclipse (TSE) white light (WL) observations. Their advantage is their spatial span starting from the chromosphere (pink emission from H$\alpha$) and extending outward into the corona (gray emission). The chromosphere is populated by spicules, dynamic jet-like structures, characterized by dense, cool neutrals, best observed in H-$\alpha$ emission, and emission from low ionized ions, as evident from ground-based observations (e.g., Meudon Observatory; see Bellot Rubio et al. 2019). Since white light is the scattered photospheric light by coronal electrons tied to the field lines our measured $L_\perp$, derived from WL images, is equivalent to a correlation between neighboring field lines. Thus, the WL intensity imaged here reflects plasma density which can be considered a proxy for the turbulence associated with velocity and magnetic field fluctuations (e.g., DeForest et al. 2016, Telloni et al. 2024).

In this study, we utilize the highest spatial resolution broadband white-light TSE image to date, taken during the 2017 August 21 TSE, shown in Fig.1 (also see Boe et al 2020 for more details), with chromospheric emission readily visible off the west limb. This image, which covers the full visible light bandpass from about 400 to 700 nm, was taken with a Celestron Edge HD, FL 2350 mm, F/10 optical system with a 2900X2900 pixel frame Fuji GFX 50s camera, and a pixel size of about 664.5 km. This image was produced using an image processing technique described by Druckmüller et al. 2006. The camera acquires data in three color channels - red, green, and blue- ; and combines them to get the white light image. These three colors can also be separated as will be discussed later in Section 3.

To compute $L_\perp$ as a function of radial distance, $L_\perp$ ($r/R_\odot$-1), we carefully select 2 regions (S1 and S2 shown by the dashed lines in Fig. 1B and C) that have clear chromospheric emission followed by coronal emission while avoiding prominences.

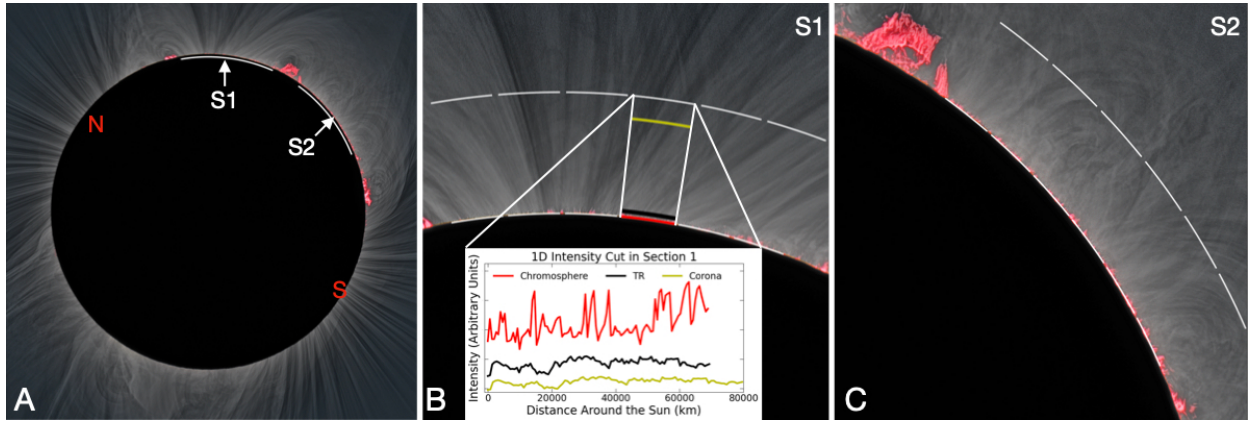

Fig. 1. White light image taken during the 21 August 2017 TSE. (A) Overview of the full eclipse image. S1 and S2 point to the regions of strong chromospheric emission and are the locations closest to the solar surface in these images. (B) and (C) are close-up views of the two sections studied in this work. We start at the base of the chromosphere and move toward the outer dashed white lines above the solar limb. We divide both S1 and S2 into six sections shown by the dashed white lines to measure $L_\perp$ twelve times at each given radius. Also shown in panel (B) is a sample plot that shows pixel intensity (in arbitrary units) versus distance for three one-dimensional cuts at different radial distances in the chromosphere (red), transition region (black), and the corona (yellow).

To estimate $L_\perp$, we first calculate the correlation function $R(\ell)$ from the white light intensity along a one-dimensional cut through the TSE image, such as the ones shown in the inset of panel (B) in Fig. 1., defined as

$$R(\ell) = \langle I(s).I(s+\ell)\rangle.$$

Here, $\langle \cdots \rangle$ is a spatial average over the total length of a selected segment. $I(s)$ is the intensity of a pixel in the white-light image at an initial coordinate $s$ along the section and $I(s+\ell)$ is the intensity at a different point at distance $\ell$. To evaluate the correlation function, we use the standard Blackman-Tukey method (Blackman & Tukey 1958, Matthaeus & Goldstein 1982) with subtraction of the local mean. After calculating $R(\ell)$ for several values of the lag $\ell$ we fit an exponential curve $R(\ell)/R(0) \sim e^{-\ell/L_\perp}$ to estimate $L_\perp$ (See Fig. 2). This analysis is repeated 12 times at each radial distance. The individual $L_\perp$-values are then averaged to obtain an estimate of $L_\perp$ at a specific radial distance away from the solar surface.

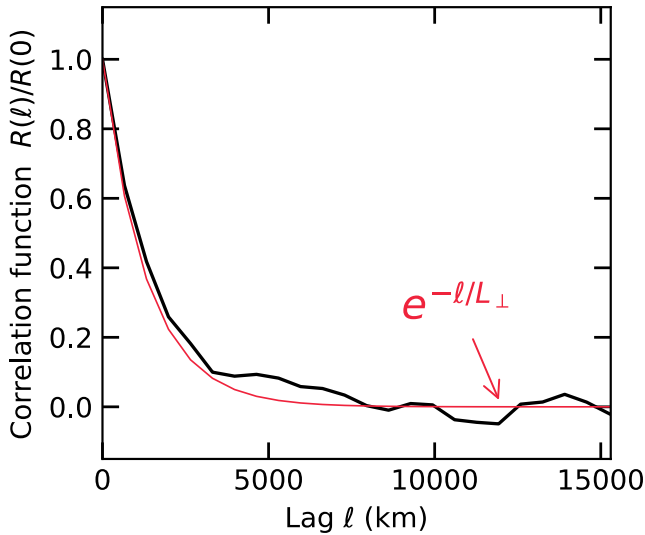

Fig. 2. Correlation function derived from TSE white light image. An example of normalized correlation function $R(\ell)/R(0)$ (black solid line) versus spatial lag ($\ell$), computed from the TSE white light image. The thin red line is an exponential fit $e^{-\ell/L_\perp}$ to the correlation function, with $L_\perp = 1325$ km.

For each section S1 and S2 shown in Fig. 1, we measure the intensities of the pixels at a height of one pixel starting from the chromosphere as a function of distance along the section S1 or S2. Effectively, this is equivalent to flying a virtual spacecraft through the image. We then step out by one pixel and repeat this process for a total of three heights. We start at the top of the chromosphere and into the corona and take the radial average of four pixels at a given latitude and continue this until we reach the end of S1 or S2. We then repeat this process for the next four consecutive pixels, until we reach a height of ~100 Mm. At this point, we have reached the tops of the coronal loops, and the structures in the image are no longer semi-radial.

We then calculate $L_\perp$ using each one-dimensional dataset. We split each S1 and S2 section into six equally sized segments, leaving us with a total of 12 datasets at each given height. Once we calculate $L_\perp$ for each of the 12 data sets at a given height, we average these measurements and assign the values to the median height for each data set.

Figure 3 is a log-log plot of $L_\perp$ as a function of radial distance $H$ in units of ($r/R_\odot$-1) from the solar surface. Based on the eclipse image, it is apparent that the first three red squares are the values of $L_\perp$ in the chromosphere starting at 0.33 Mm, which is roughly constant with radial distance. The following blue round symbols are $L_\perp$ directly above the chromosphere and into the TR and inner corona. We note that the first blue circle, at $H \approx 5.7 \times 10^{-3}$ $R_\odot$ (4 Mm), follows the same trend as the chromospheric values. This point is then followed by two that rise very sharply, up to $1.3 \times 10^{-2}$ $R_\odot$ (9.3 Mm). The rest of the measurements rise much more slowly, ending at 0.135 $R_\odot$ (94.4 Mm). We also include a magenta triangle for the only other inference of $L_\perp$ in the inner corona from the previous work of Sharma & Morton (2023).

On the left side of the plot, a red cross marks the typical size of a solar granule in the photosphere (Schwarzschild 1959, Title et al. 1988, Rimmele et al. 2020, Wöger et al. 2021). The measured $L_\perp$ values in the chromosphere are very close to the average solar granulation width (~ 1.5 Mm). Although we do not observe the solar photosphere directly, this result is a strong indication that the solar granules drive the turbulence near the solar surface and determine its injection scale.

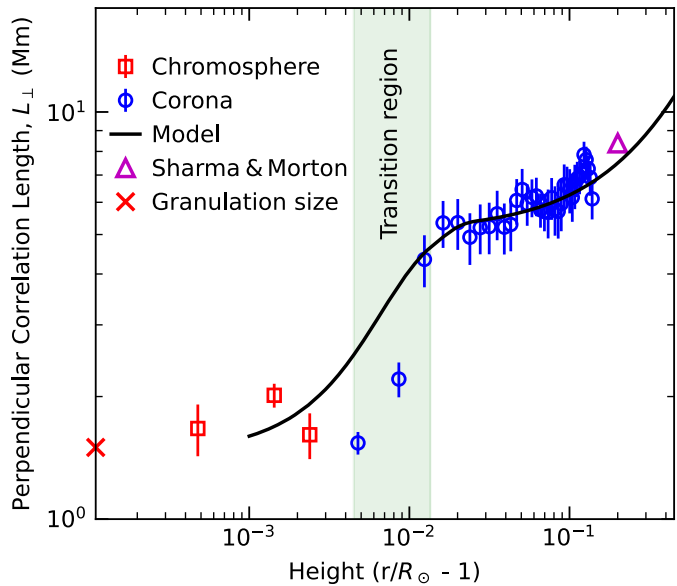

Fig. 3. Variation of perpendicular correlation length as a function of height above the solar surface, given in units of a solar radius ($R_\odot \approx 700$ Mm). The red squares are measurements in the chromosphere and blue round symbols are measurements in the transition region and the lower corona from this work, using white light eclipse image. Previous closest estimate of $L_\perp$ in the corona (Sharma & Morton 2023) is shown by a magenta triangle. A scaled version of the theoretical model of $L_\perp$ (Cranmer & van Ballegooijen 2012), based on the equation $\boldsymbol{L_\perp \alpha\, B_0^{-1/2}}$ (Hollweg 1986) is plotted in black solid line. The typical width of solar granules is (~1.5 Mm) shown by a red cross at the left edge.

We further compare the observational measurements with a theoretical estimate of the radial evolution of $L_\perp$. We used the work by Cranmer & van Ballegooijen (2012) which is based on the proportionality $L_\perp \alpha\, B_0^{-1/2}$ derived in Hollweg (1986), with the magnetic field given by Banaszkiewicz et al. (1998). We refer to this model (Hollweg 1986 and Cranmer & van Ballegooijen 2012) as the HC model. Their best determined $L_\perp$ at the photosphere was set at 0.12 Mm and was the starting point for the rest of the model (Cranmer & Ballegooijen 2005, Cranmer 2007). The HC model is shown in Figure 3 as a solid black curve. However, for the model to fit our $L_\perp$ measurements we set the initial photospheric value to 1.5 Mm. (Note that a similar photospheric value of $L_\perp$ was used by Sharma & Morton (2023) to match the only coronal data in that work). Surprisingly, our ~ 40 measurements follow this scaled model strikingly well from 0.33 Mm to ~ 94 Mm, filling the missing data below the Sharma & Morton (2023) data point. This reaffirms that the $L_\perp$ value in the vicinity of the photosphere is ~ 1.5 Mm, which is close to the granulation size.

## 3. Empirical Inference of the Height of the Transition Region

The way our measurements differ from the HC model is in the rapid rise of $L_\perp$ between 4 and 9 Mm above the solar surface (green shaded region in Fig. 3). Since the HC model assumes that $L_\perp \alpha\ B_0^{-1/2}$ and the conservation of magnetic flux implies the expansion of the magnetic flux tube is inversely proportional to its cross-sectional area, we can use the change in $L_\perp$ to infer the expansion of the magnetic field (derived below).

$$B \propto L_\perp^{\frac{1}{2}}$$

$$A \cdot B = const$$

$$L_\perp \propto A^{-\frac{1}{2}}$$

where A is the cross section of a magnetic flux tube. With this assumption, we can use our measurements of $L_\perp$ (Fig. 3) to produce an empirical description of the expansion of the magnetic field starting from 0.33 Mm above the solar surface (see Fig. 4). We see that the magnetic field starts with little to no expansion from the chromosphere until 4 Mm. This is followed by a rapid expansion until 9 Mm, and then a shallower expansion from 9 Mm until ~100 Mm into the corona. We connect this rapid expansion of the magnetic field with the TR and place it at a height of ~ 4 Mm above the photosphere. Our empirical value is consistent with previous inferences (e.g., Horne et al. (1981), Anderson & Athay (1989), Belkora et al. (1992), Fontenla et al. (1993), Ewell et al. (1994), Johannesson & Zirin (1996), Zirin (1996), Zhang et al. (1998), Song P. (2023)).

However, our empirical estimate of the thickness of the TR of about 5 Mm is significantly different from published values of 100 – 200 km (e.g., Gabriel 1976, Fontenla et al. 1993, Avrett & Loeser 2008, Cranmer 2007, Song et al. 2023). While estimates of the transition region thickness are based mainly on models of the temperature and density, our empirical results suggest that the transition region is rather tied to the rapid expansion of the magnetic field between the chromosphere and the corona as shown in Fig. 4.

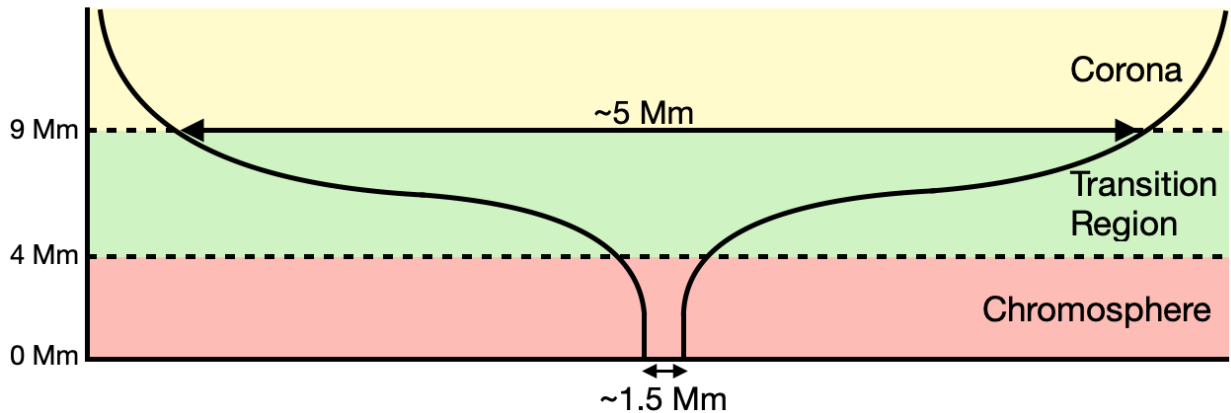

Fig. 4. Schematic diagram (not to scale) of the expansion of the magnetic field, starting from the chromosphere into the corona as inferred from the $L_\perp$ measurements. This is similar to the schematics by Rimmele et al. (2020) and Wöger et al. (2021), except for the heights which are different.

To further confirm that the labels in Fig. 4- chromosphere, TR, and corona- are supported by the data, we take advantage of the three colors (red, green, and blue) that can be separated in the broadband WL images. To distinguish between the contribution from each color a small region in S2 was selected (~0.1 rad). The obtained intensities were averaged in all three channels, as well as the WL intensities, and then plotted as colored symbols in Fig. 5. The green shaded region corresponds to the same TR identified in Fig. 3. To the left of the green shaded region the WL intensities are dominated by the red channel which is expected because the chromosphere is dominated by H-$\alpha$ emission along with other emission lines such as the Fe X red line at 637.4 nm. In the TR the intensity of the three colors start to become comparable and in the coronal region all three intensities are practically the same.

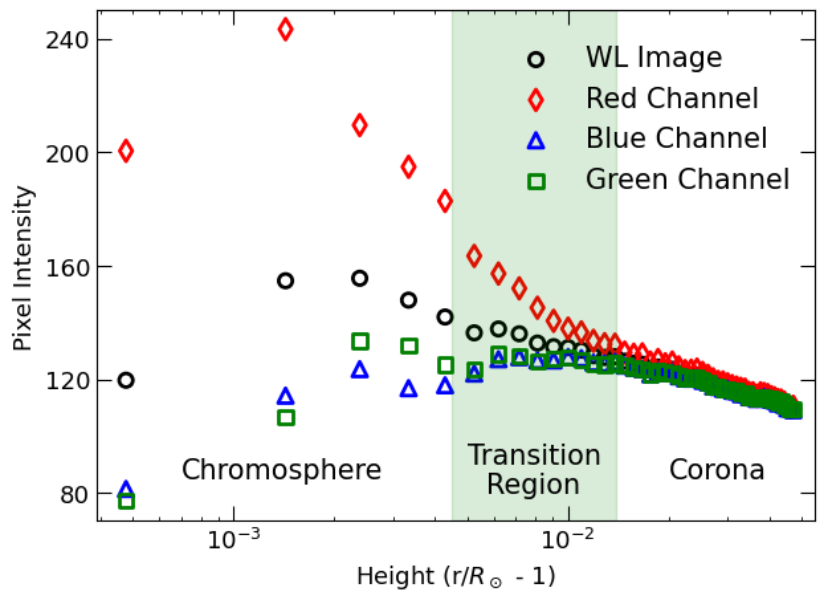

Fig. 5. Pixel intensities in the different color channels of the detector versus height.

## 4. Summary and Conclusion

In this study, we present the first empirical measurements of turbulence injection scales in the solar chromosphere and transition region, using a high-resolution white-light image of the solar atmosphere from the August 21, 2017 total solar eclipse. This image made it possible to explore turbulent structures starting from the chromosphere into the lower corona at unprecedented spatial scales. While line of sight integration effects are inevitable in all remote sensing observations, the systematic variation of the correlation length with height suggests that such effects are minimal here.

Although we do not observe the photosphere in this work, disturbances in the magnetic fields emerging from the photosphere are likely drivers of turbulence in the chromosphere. Therefore, one expects that the turbulence in the chromosphere will bear at least some signature of the photospheric dynamics. Since the measured $L_\perp$ values become saturated ~ 1.5 Mm - a value close to the granules-scale size- our interpretation suggests that granules play a key role in setting the driving scale of turbulence near the photosphere. Sharma and Morton (2023) found "that the corona shows fluctuations (which) are coherent over scales of $L_\perp \approx 7.6-9.3$ Mm perpendicular to the magnetic field, which is indicative of supergranulation scales." However, interpreting the large correlation length in the corona as the injection scale at the 'coronal base' is not supported by this study. Rather, our results which extend closer to the solar surface show that the original energy injection at the base happens on a much smaller scale. Therefore, our finding suggests that solar granules, rather than supergranules, play a key role in driving the turbulent energy injection into the chromosphere. However, modeling efforts are required to draw a definite conclusion.

The results of this paper provide crucial insights for solar wind modeling, especially in regions close to the Sun. Coronal heating models often use the value of $L_\perp$ as a critical input due to its relationship with turbulent heating (e.g., Matthaeus et al. 1999). Recent coronal heating models have started including the chromosphere, i.e. starting at levels lower than the transition region (e.g., Usmanov et al. 2018). However, current models rely on assumed, ad-hoc values for the turbulence outer scale at the solar wind source, with estimates spanning from the supergranulation scale to the magnetic flux rope size (e.g., Cranmer & van Ballegooijen 2012). Our results suggest a need for substantial revisions in such models.

Our data further provide observational insights into the height of the transition region, placing it around 4 Mm above the solar surface. We observe a rapid increase in $L_\perp$ within this region, reflecting a sharp expansion in magnetic field lines, associated with the formation of the transition region. Previous estimates from models (e.g., Horne et al. (1981), Anderson & Athay (1989), Belkora et al. (1992), Fontenla et al. (1993), Ewell et al. (1994), Johannesson & Zirin (1996), Zirin (1996), Zhang et al. (1998), Song P. (2023)) placed the TR height between 2 and 8 Mm, which is consistent with our results.

However, our inference of the thickness of the transition region of about 5 Mm is in strong contrast with previous inferences from modeled temperatures and densities which yield a value of ~100 – 200 km (e.g., Gabriel 1976, Fontenla et al. 1993, Avrett & Loeser 2008, Cranmer 2007, Song et al. 2023). Our comprehensive results provide a new approach; they not only yield an empirical estimation of the height of the transition region but also spatially resolve its steep gradient (Mariska 1992).

## Acknowledgments

We thank Prof. Steven Cranmer for providing the data for the theoretical model curve. ZB and SH were supported by National Science Foundation- Division of Atmospheric and Geospace Sciences grant 2313853 to the University of Hawaii. RB was supported by National Aeronautics and Space Administration grant 80NSSC21K1767. We also thank the two anonymous reviewers for their insightful comments on this manuscript.